\documentclass[namedreferences,hyperref,optionalrh]{spr-sola}
\usepackage{graphicx}        
\usepackage{color}           
\usepackage{lmodern}
\usepackage{amsmath}
\usepackage{comment}

\chardef\us=`\_

\begin{document}

\begin{frontmatter}
\title{Magnetic Reconnection at Hyperbolic Flux Tube associated with a Confined Flare in NOAA Active Region 12268}

\author[addressref={aff1}]
{\inits{P.}\fnm{Pawan}~\snm{Kumar}\orcid{0009-0000-3324-2987}}

\author[addressref={aff1}]{\inits{Sadashiv}\fnm{Sadashiv}\orcid{0009-0003-0045-4387}}

\author[addressref={aff1},email={sainisanjay35@gmail.com}]{\inits{Corresponding author: S.}\fnm{Sanjay}~\snm{Kumar}\orcid{0000-0003-4578-6572}}

\author[addressref={aff2,aff3}]{\inits{S.S.N.}\fnm{Sushree S.}~\snm{Nayak}\orcid{0000-0002-4241-627X
}}

\author[addressref=aff4]{\inits{S.K.}\fnm{Simrat}~\snm{Kaur}\orcid{0009-0003-5326-9310}}


\author[addressref=aff4]{\inits{R.B.}\fnm{Ramit}~\snm{Bhattacharya}\orcid{0000-0003-4522-5070}}

\address[id=aff1]{Department of Physics, Patna University, Patna 800005, India}

\address[id=aff2]{Department de F\'isica, Universitat de les illes Balears, E-07122, Palma de Mallorca, Spain}

\address[id=aff3]{Institut d'Aplicacions Computacionals de Codi Comunitari (IAC3), Universitat de les illes Balears, E-07122, Palma de Mallorca, Spain}

\address[id=aff4]{Udaipur Solar Observatory, Physical Research Laboratory, Dewali, Bari Road, Udaipur
313001, India}



\runningauthor{Kumar et al.}
\runningtitle{Reconnection at HFT associated with a confined flare }

\begin{abstract}
In this paper, we identify the magnetic reconnections at the hyperbolic flux tube (HFT), aided by slipping reconnection at quasi-separatrix layers (QSLs), which are pivotal to the occurrence of a confined M2.1 class flare in NOAA active region 12268.  The magnetic field topology before the flare's onset is obtained through a non-force-free-field extrapolation scheme that accommodates a non-zero Lorentz force. A key aspect is the presence of an HFT in the computational domain above the flaring region, along with two QSLs at the lower boundary.
To simulate the dynamics of the active region, we conduct a data-constrained magnetohydrodynamics (MHD) simulation initiated by the extrapolated field. The dynamics captured in the simulation document the formation of a current sheet within the HFT configuration, leading to magnetic reconnection at the HFT. Additionally, we observe the slipping motion of the footpoints of the magnetic field lines in the QSLs at the bottom boundary, which indicates the occurrence of slipping reconnection in the QSLs. Importantly, the magnetic reconnection at the HFT is suggested to be the primary driver in the development of the intricate flare brightenings and the flare ribbons. 

\end{abstract}
\end{frontmatter}

\section{Introduction}
     \label{S-Introduction}
Solar flares are explosive phenomena observed in the solar atmosphere, resulting in intensity peaks along a wide range of electromagnetic waves such as radio, visible light, X-ray, and gamma rays \citep{2011LRSP....8....6S}. Magnetic reconnection is believed to be one of the mechanisms behind flares, which converts magnetic energy into other forms, such as heat, kinetic energy of mass flow, and acceleration of the charged particles \citep {2011LRSP....8....6S,2023AdSpR..71.1856P}.

Some flares are associated with coronal mass ejection (CME), also known as `eruptive flares' \citep{2001ApJ...559..452Z,2008ApJ...673L..95T}. The eruptive flares are generally explained by the standard solar flare model, a.k.a CHSKP model \citep{1964NASSP..50..451C,1966Natur.211..695S,1974SoPh...34..323H,1976SoPh...50...85K}. The model proposes that a flare is driven by the sudden rise of a pre-existing magnetic flux rope (bundle of twisted field lines wrapping around an axis). This leads to magnetic reconnection between the legs of magnetic field lines stretched out by the rising flux rope. The energy released during the reconnection causes the parallel flare ribbons. However, various studies depict flares occurring differently than suggested by the standard flare model, with ribbons forming in a variety of geometries \citep{1996SoPh..168..115M,2000ApJ...540.1126A,2009ApJ...700..559M,2012ApJ...760..101W,2019ApJ...881..151L,2020SoPh..295...75D}. 

The flares that do not have an association with a CME are termed confined flares.  In the context of explaining confined flares, the constraining effect of the background magnetic field that lies above the flaring region is regarded as a key factor \citep{2007ApJ...665.1428W, 2014ApJ...793L..28Y, 2017ApJ...843L...9W}. The nonpotentiality of active regions (ARs) responsible for producing the flares is another crucial factor that helps to determine whether solar flares are confined or eruptive \citep{2004ApJ...616L.175N, 2015ApJ...804L..28S}.
Recently, \citet{2019ApJ...881..151L, 2022ApJ...933..191D} carried out statistical studies of  confined flares and found that there are two types of confined flares.
``Type I" confined flares are characterized by slipping reconnection in a complex magnetic configuration with a stable filament. In contrast, ``Type II" flares are associated with failed eruptions of the filaments due to a strong strapping field, which can be explained by the standard flare model.
  However, a comprehensive understanding of magnetic reconnection underlying confined flares in complex magnetic topology demands further investigation that involves the evolution of coronal magnetic configurations, which contribute to magnetic reconnection.  In recent times, several studies have been conducted using magnetohydrodynamics (MHD) simulations to explain the onset of the confined flares with complex ribbon geometry \citep{2018ApJ...860...96P,2019ApJ...875...10N,2024SoPh..299...15A,2025SoPh..300...79N,2025SoPh..300...22K}. In these studies, magnetic features such as three-dimensional (3D) null points, quasi-separatrix layers (QSLs), and hyperbolic flux tubes (HFTs) are identified as the potential sites of magnetic reconnection. Relevantly, QSLs are regions of heightened magnetic field gradient with distortion in field line mapping, while HFT is the intersection of two QSLs \citep{1999A&A...351..707T,2002JGRA..107.1164T}. Reconnection at QSLs has been shown to initiate the flare in the work of \citet{2018IAUS..340...81P,2021SoPh..296...26K,2025SoPh..300...22K}, while several studies have reported reconnection at HFT behind the flares \citep{2017A&A...604A..76M,2021MNRAS.503.1017M,2024SoPh..299...15A}. Notably,  \citet{2006SoPh..238..347A} attributed the rapid slippage of field lines in QSLs to a confined flare in the absence of a flux rope.

This work focuses on explaining the onset and the associated observed features of a confined M2.1 class flare that occurred on 2015 January 29 in the NOAA AR 12268. Notably, earlier investigations into this flaring event were performed by \citet{2017ApJ...847..124H} and \citet{2019ApJ...871..105Z}, who utilized observations and numerical modeling through the non-linear-force-free-field (NLFFF) extrapolation \citep{2008JGRA..113.3S02W}. \cite{2017ApJ...847..124H} reported the development of two primary ribbons at the main flare site, followed by the appearance of two secondary ribbons at remote locations. They suggested that the primary ribbons were caused by magnetic reconnection in a fan-spine configuration, while the formation of secondary ribbons was attributed to the dissipation of kinetic energy from the plasma flow (heating due to compression). On the other hand, \cite{2019ApJ...871..105Z} proposed an alternative mechanism, asserting that reconnections in a sheared arcade, a null point, and QSLs are essential to the formation of the observed ribbons. To delve deeper into the magnetic reconnection process that underlies the flaring event, we adopt the advanced methodology of data-constrained MHD simulation, which is initiated from an extrapolated magnetic field. For constructing the extrapolated magnetic field, we utilize a non-force-free-field (NFFF) extrapolation which supports the non-zero Lorentz force \citep{2008ApJ...679..848H,2010JASTP..72..219H}. The force is crucial in autonomous generation of the simulated dynamics {\citep{2018ApJ...860...96P}}. In the extrapolated magnetic configuration, we identify a HFT in the flaring region along with the QSLs at the bottom boundary.  The simulated dynamics documents the formation of current sheets in the HFT region, leading to magnetic reconnection at the HFT. Additionally, the simulation reports the slipping  motion of field line footpoints in the QSLs --- signaling toward the role of slipping reconnection in the onset of the flare \citep{2006SoPh..238..347A}. The present simulation suggests that reconnection at the HFT, aided by slipping reconnection at the QSLs, is central to produce the different observational features, including the complex ribbon morphology of the flare.

This work is presented as follows: in Section \ref{Observation and Extrapolation}, we provide the key observational features of the flare and the NFFF extrapolated field line topology of the active region. In section \ref{MHD-simulation}, we briefly describe our simulation scheme, including the numerical model and computational setup. Section \ref{Results of MHD Simulation} shows the simulation results, and finally, Section \ref{Summary and Discussion} summarizes and discusses the key findings of the work.

\section{Observations and Extrapolation}
\label{Observation and Extrapolation}
\subsection{Multi-wavelength Observations of the M2.1 Class Flare }
\begin{figure}
    \centering
    \includegraphics[width=1\linewidth]{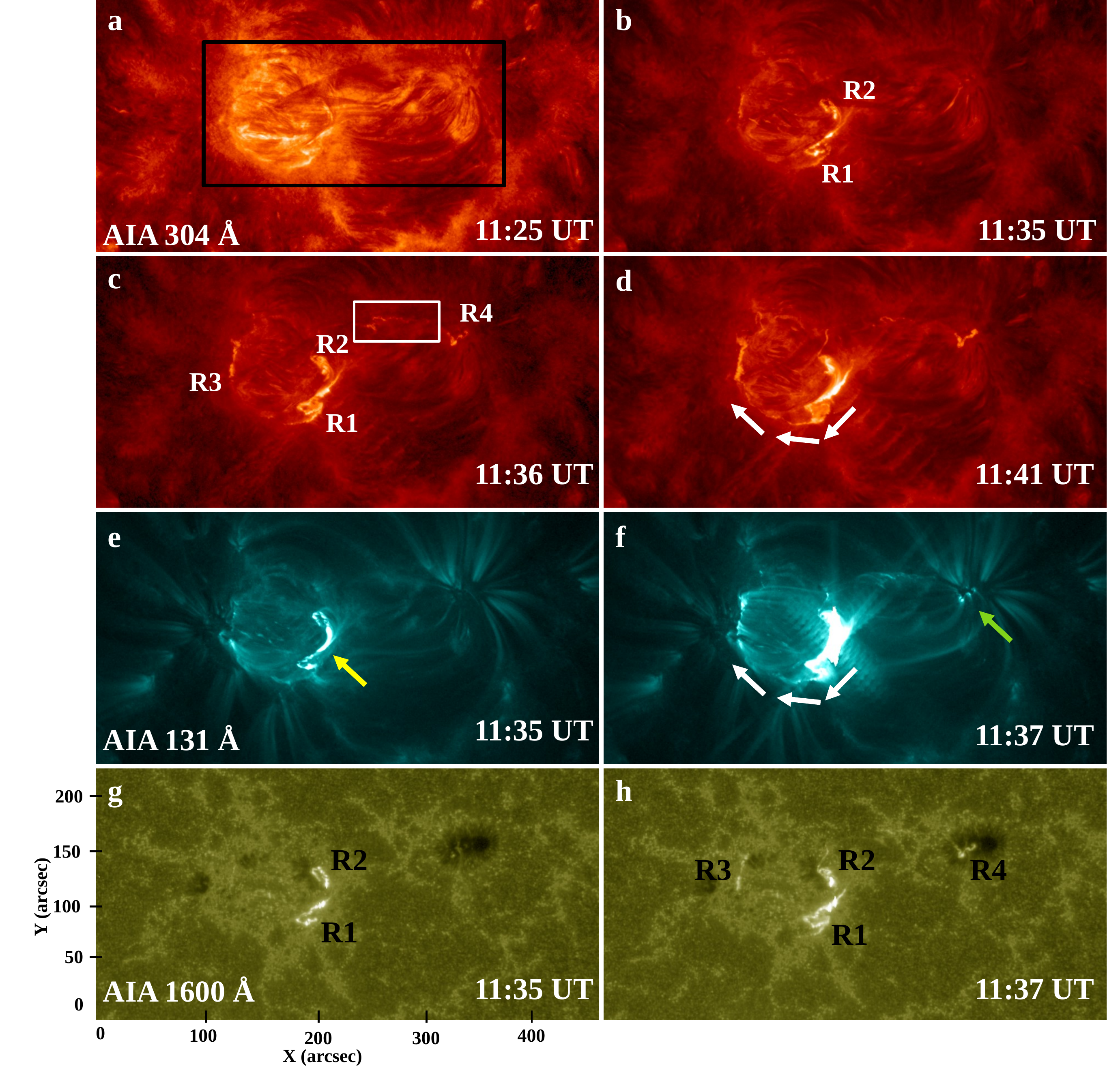}
    \caption{The flare observations in AIA 304 $\text{\AA}$ (panels (a)-(d)), 131 $\text{\AA}$ (panels (e)-(f)) and 1600 $\text{\AA}$ (panels (g)-(h)). The black rectangle in panel (a) encloses the flaring region. Panel (b) depicts the primary ribbons R1 and R2 constituting the central brightening observed just after the flare onset. Panel (c) documents the appearance of the secondary ribbons, R3 and R4, with an additional faint brightening trace (highlighted by a white rectangle). Panel (d) displays the extension of the central brightening in a clockwise direction (shown by white arrows) --- resulting in a faint quasi-circular bright structure. Panel (e) shows the central bright region depicted by a yellow arrow.  
    A faint quasi-circular bright structure (denoted by white arrows) and a remote brightening (marked by a green arrow) are also visible in 131 $\text{\AA}$ (panel (f)).   
  Panel (g) identifies the appearance of two primary flare ribbons (denoted by R1 and R2) in 1600 $\text{\AA}$ in the central region. Panel (h) displays the generation of two more remote ribbons, as denoted by R3 and R4.}
    \label{fig1}
\end{figure}

The M2.1 class flare occurred in NOAA AR 12268 on 2015 January 29. It initiated at around 11:32 UT, reached its peak at around 11:42 UT as reported in the GOES observations. Multi-wavelength observational study of this event has already been done by \cite{2017ApJ...847..124H} and \cite{2019ApJ...871..105Z}. Figure \ref{fig1} depicts its key observational features in different channels of the Atmospheric Imaging Assembly \citep[AIA,][]{2012SoPh..275...17L} onboard the Solar Dynamic Observatory 
\citep[SDO,][]{2012SoPh..275....3P} which observes the Sun in EUV and UV wavelengths with a pixel size of 0.6$''$ and with a time cadence of 12 seconds (for EUV) and 24 seconds (for UV).
Panels (a) to (d) of Figure \ref{fig1} show its observational features in AIA 304 $\text{\AA}$. The flaring region is highlighted by a black rectangle in panel (a). A few minutes after the flare initiation, we observe two primary flare ribbons R1 and R2, constituting the brightening in the central region (panel (b)). In panel (c), two secondary ribbons, R3 and R4, appear on the left and right sides of the primary ribbons, respectively. In addition to that, we notice a presence of faint brightening trace (shown in the white rectangle in panel (c)). Further, as the time advances towards the peak of the flare, the central brightening seems to extend in the clockwise direction to form a faint quasi-circular brightening (as illustrated by white arrows in panel (d)). Panel (e) illustrates the central brightening indicated by a yellow arrow in AIA 131 $\text{\AA}$, while panel (f) displays the existence of the quasi-circular structure that arises from the expansion of the central brightening (denoted by white arrows) and the distant brightening (identified by a green arrow). Panel (g) illustrates the appearance of two primary ribbons, denoted by R1 and R2, in AIA 1600 $\text{\AA}$. After that, we 
also observe the formation of two secondary ribbons (highlighted by R3 and R4 in panel (h)) located on left and right of the primary ribbons. The secondary ribbons are faint in comparison to the primary ribbons. Relevantly, \citet{2017ApJ...847..124H} have reported the evolution of the R1 and R2 ribbons in a quasi-parallel manner. Moreover, they have also suggested the ribbons R1 and R3 to be the parts of the same quasi-circular ribbon that extends from R1 to R3.

\subsection{Non-Force-Free-Field Extrapolation of AR 12268 }
\label{nfff}

To obtain the initial magnetic field topology of the NOAA AR 12268, we perform the extrapolation using a non-force-free-field model  \citep{2008ApJ...679..848H,2010JASTP..72..219H}. This model, which is founded on the principle of minimum dissipation rate (MDR) 
\citep{2007SoPh..240...63B}, defines the magnetic field $\bf{B}$ as follows:

\begin{equation}
\label{ext field eq}
    \mathbf{B}=\mathbf{B}_{1}+\mathbf{B}_{2}+\mathbf{B}_{3} ; \nabla \times \mathbf{B}_{i}=\alpha_{i}\mathbf{B}_{i}
\end{equation}
where ${i}=1,2,3$. Evidently, the magnetic field $\bf{B}$ is a result of the superposition of three linear force-free magnetic fields. For practical purposes, without loss of generality, one of the fields is considered to be potential such that $\alpha_1\not=\alpha_2\not=0$ and $\alpha_3=0$. An iterative approach is used to find the optimal pair ($\alpha_1,\alpha_2$) by minimizing the average deviation between the observed transverse component of the field ($\mathbf{B}_{t}$) and the obtained transverse component ($\mathbf{b}_{t}$) in each iteration. The deviation is given by \citep{2010JASTP..72..219H}:
\begin{equation}
\label{ext error eq}
    E_n=\frac{\sum_{k=1}^{M} |\mathbf{B}_{t,i}-\mathbf{b}_{t,i}| \times \mathbf{B}_{t,i}}{\sum_{k=1}^{M} |\mathbf{B}_{t,i}|^2}
\end{equation}
where M represents the total number of grid points on the transverse plane. Remarkably, the recent investigations have successfully accounted for a variety of transient events, such as flares, coronal jets, and coronal dimmings, by employing MHD simulations that are initiated by NFFF extrapolation \citep{2018ApJ...860...96P,2019ApJ...875...10N,2020ApJ...903..129P,2021PhPl...28b4502N,2024ApJ...975..143N,2025SoPh..300...22K}.

For the NFFF extrapolation of AR 12268, we utilize the Helioseismic and Magnetic Imager 
 \citep[HMI,][]{2012SoPh..275..229S} magnetogram obtained from the `hmi.sharp\_cea\_720s' data series that provides photospheric vector magnetogram of the Sun, with a time resolution of 12 minutes and a pixel resolution of 0.5$''$. 
To derive magnetic fields in a Cartesian grid, the magnetogram is first remapped onto a Lambert Cylindrical Equal-area (CEA) projection, and then converted into heliographic coordinates \citep{1990SoPh..126...21G}. The magnetogram taken at 11:24 UT (8 minutes prior to the flare's commencement) is utilized for the extrapolation. The magnetogram cutout measures $912 \times 456$ pixels in the x and y directions in Cartesian coordinates, which is subsequently rescaled to $456 \times 228$ pixels due to computational restrictions. Moreover, we take the length along the z-axis to be $228$ pixels, resulting in a computational volume of $456 \times 228 \times 228$ pixels. This translates to approximately 328 Mm in the x direction and 164 Mm in both the y and z directions.

\begin{figure}
    \centering
    \includegraphics[width=1\linewidth]{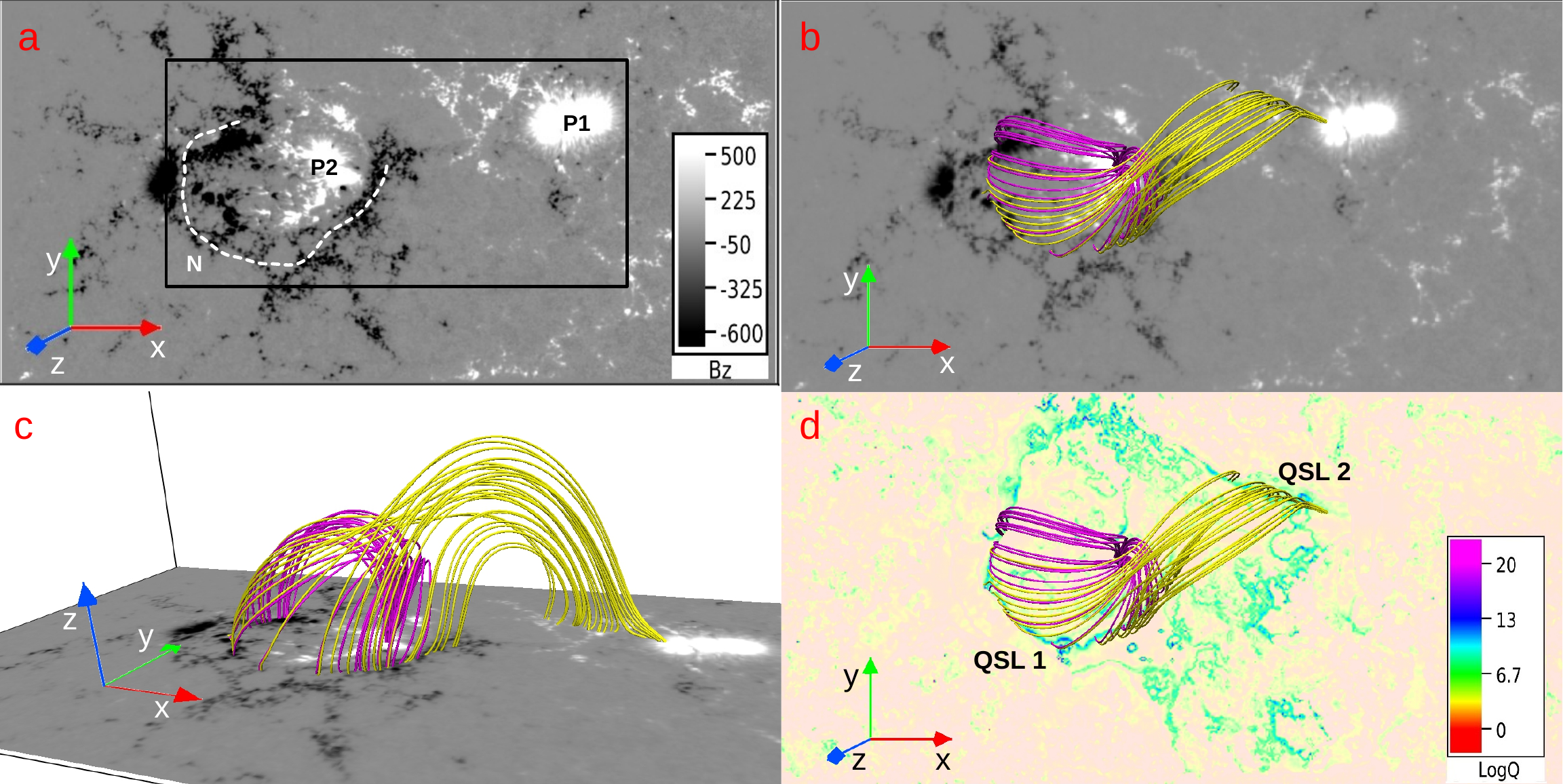}
    \caption{Panel (a) shows the LOS magnetic field ($B_z$) strength (in Gauss) at 11:24 UT in grayscale, in which the black rectangle marks the flaring region. Two major positive polarities are denoted as P1 and P2. Quasi-circular negative polarity (highlighted by a dashed curve) is marked by N.  Panels (b) and (c) show the top and side views of the extrapolated magnetic field lines overlaid with $B_z$ at the bottom boundary, respectively. Panel (d) plots the value of LogQ at the bottom boundary, superimposed with the field lines. }
    \label{fig2}
\end{figure}

\begin{figure}
    \centering
    \includegraphics[width=1\linewidth]{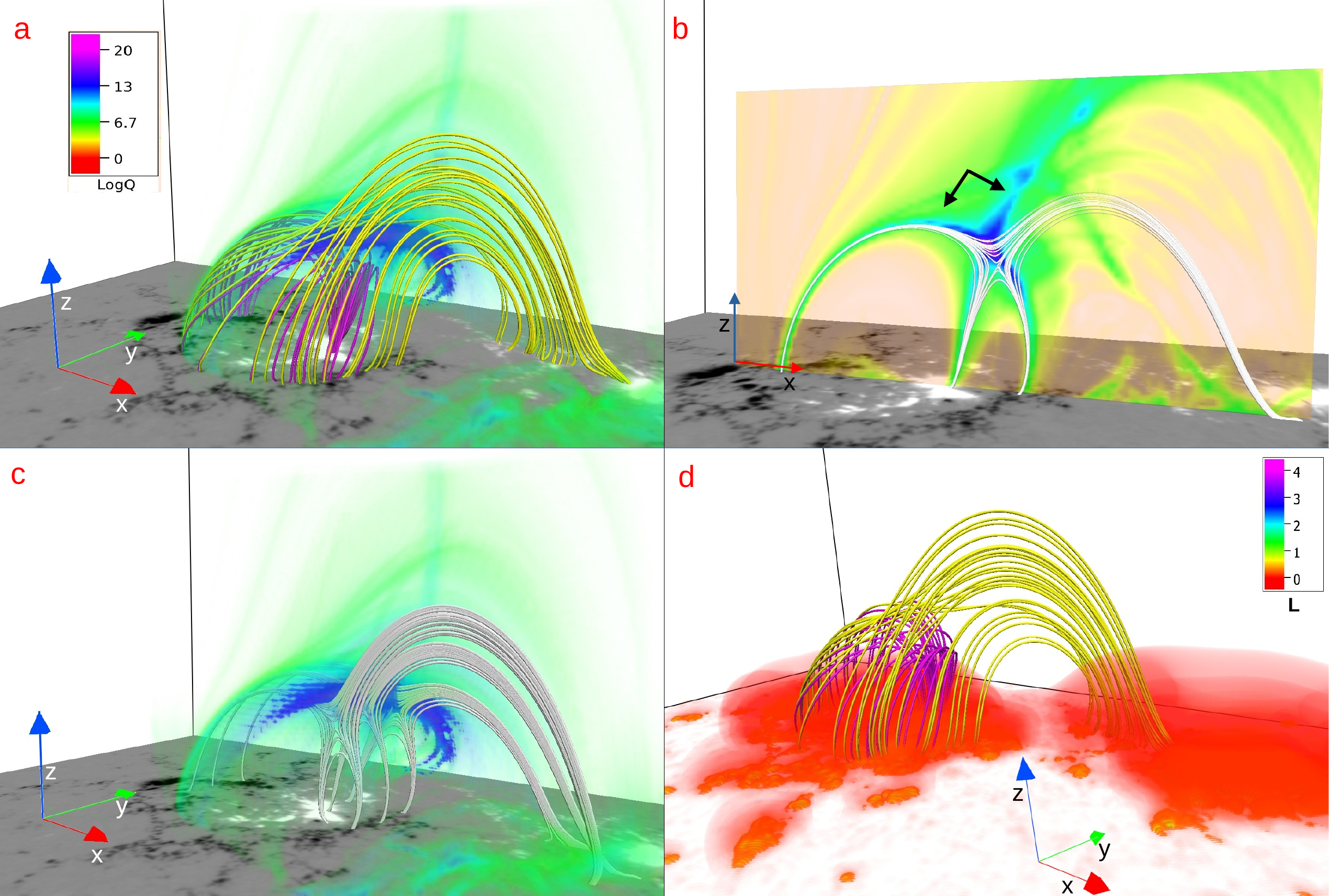}
    \caption{Panel (a) displays the extrapolated field lines along with the direct volume rendering of LogQ. Panel (b) plots the projected field lines on a $y$-constant plane superimposed with LogQ, where the black arrows highlight an intersection of two QSLs. 
    Panel (c) shows the project field lines at three different $y$-constant planes with LogQ. 
 Panel (d) plots the distribution of the Lorentz force density strength (L) with the extrapolated field lines.}
    \label{fig3}
\end{figure}

Figure \ref{fig2}(a) presents the observed line-of-sight (LOS) magnetic field strength depicted in grayscale. It is significant to observe the existence of two major positive-polarity regions, identified as P1 and P2, where the left region (P2) is encircled by negative polarity in a quasi-circular arrangement, denoted by N.  Panels (b) and (c) of Figure \ref{fig2} illustrate the top and side views of the extrapolated field lines in the flaring region, respectively. 
Here it is important to mention that we have chosen the seed points of the extrapolated field lines in the vicinity of the HFT, the location of which is shown in Figure \ref{fig3}.
The extrapolated field lines reveal two separate magnetic configurations arising from different connectivity domains. The first configuration is generated from the connectivity between polarities P2 and N (shown by pink field lines), with footpoints forming a quasi-circular shape in the negative polarity. The second configuration is represented by yellow field lines, which connect the positive plage polarity (located east of P1) to the negative polarity N. The footpoints of these field lines trace a quasi-circular shape in the N region. In Figure \ref{fig2}(d), we superimpose the extrapolated field lines with the squashing factor \citep[Q,][]{2002JGRA..107.1164T, 2016ApJ...818..148L} at the lower boundary. It is significant to note that high Q values (LogQ $>$ 8) are present near the footpoints of the field lines at two different locations, suggesting the existence of two QSLs at the bottom boundary, labeled as QSL1 and QSL2 in the figure.

The distribution of Q-values within the computational volume is quite interesting, as shown in Figure \ref{fig3}(a)-(c). In particular, panel (a) shows the presence of high Q-values in the region where the two configurations are in close proximity. To further explore, panel (b) displays the Q-values in a $y$-constant plane located in the region. This panel is further overlaid with the field lines of the transverse field in white (achieved by setting $B_y=0$ in $\mathbf{B}$), which represent the projected magnetic field lines on the $y$-constant plane. The intersection of the two QSLs is evident, along with the presence of X-shaped field lines of the transverse field in the vicinity of the QSLs (as denoted by black arrows in panel (b)). Such a convergence of QSLs with the X-shaped arrangement of the projected field lines can be identified as HFT \citep{2002JGRA..107.1164T, 2021SoPh..296...26K}. It is evident that the alignment of the field lines around the HFT is anticipated to be conducive to the initiation of magnetic reconnection.
Additionally, the intersection of the two QSLs traces an elongated shape when plotted in the volume, and a series of X-shaped geometries of the transverse field is obtained along this elongated shape (see panel (c)). It is noteworthy that the overall magnetic field configuration resembles the fan-spine structure of a 3D null. To find magnetic nulls in the flaring region, we employed the trilinear method of null detection \citep{Haynes_2007,2020A&A...644A.150O} for the NFFF extrapolated field, which confirmed a complete absence of nulls in the flaring region.

In Figure \ref{fig3}(d), we illustrate the direct volume rendering of the magnitude of Lorentz force density with the extrapolated field lines. It is noteworthy that the force is mostly present at lower heights and nearly diminishes at larger heights, indicating that the corona approaches a near force-free equilibrium \citep{2021LRSP...18....1W}. Furthermore, at the lower heights, the force predominantly exists in the flaring region and is essential for the self-consistent generation of the simulated dynamics of the active region, ultimately leading to the initiation of the flaring event.


\section{MHD Simulation of AR 12268}
\label{MHD-simulation}
\subsection{Governing Equation and Numerical Model} 
To understand the onset of the M2.1 class flare, we simulate the evolution of the AR 12268 by performing a data-constrained MHD simulation that is initiated with the NFFF extrapolated field.
For the purpose, the plasma is considered  to be incompressible, thermally homogeneous, and perfectly electrical conducting \citep{2017PhPl...24h2902K, 2020ApJ...892...44N}.
This consideration is adequate for examining the initiation mechanism of coronal transients, especially the alterations in field line connections resulting from reconnection, as demonstrated in previous studies \citep{2018ApJ...860...96P, 2019ApJ...875...10N, 2020ApJ...903..129P, Bora_2023, 2025SoPh..300...22K}. The resultant MHD equations in their dimensionless form, are given as:

\begin{equation}
    \label{momentum transport}
    \frac{\partial \mathbf{v}}{\partial t}+(\mathbf{v}\cdot\nabla)\mathbf{v} = -\nabla {p} + (\nabla \times \mathbf{B})\times\mathbf{B}+\frac{\tau_\text{a}}{\tau_\nu}\nabla^2\mathbf{v},
\end{equation}
\begin{equation}
    \label{continuity equation}
    \nabla.\mathbf{v}=0,
\end{equation}
\begin{equation}
    \label{induction equation}
    \frac{\partial \mathbf{B}}{\partial t}=\nabla \times\mathbf{v}\times\mathbf{B},
\end{equation}
\begin{equation}
    \label{solenoidal condition}
    \nabla .\mathbf{B}=0, 
\end{equation}

\noindent The magnetic field $\bf{B}$ and plasma velocity $\bf{v}$ in these equations are normalised with average field strength ${B}_{0}$ and Alfven speed $\nu_{a}=\frac{{B}_{0}}{\sqrt{{4}\pi\rho_{0}}}$ respectively, with $\rho_{0}$ being the constant mass density. Further, the length scale ${L}$ and time scale ${t}$ are normalised with the length scale of the vector magnetogram ${L}_{0}$ and Alfvenic transit time $\tau_{a}={{L}_{0}}/{\nu_{a}}$ respectively. The viscous diffusion time scale is $\tau_{v}={L}_{0}^2/\nu$, where $\nu$ is the coefficient of kinematic viscosity. The plasma pressure $p$ is scaled with $\rho\nu_{a}^2$, which follows the equation given as:

\begin{equation}
     \label{pressure equation}
    \nabla^2({p}+\frac{\text{v}^2}{2})=\nabla\cdot[(\nabla\times\mathbf{B})\times\mathbf{B}-(\nabla\times \bf{v})\times \mathbf{v}].
\end{equation}

This equation is formulated by applying equation \ref{continuity equation} to equation \ref{momentum transport}. It is apparent that the pressure adjusts itself instantaneously to maintain the incompressible nature of the flow, which results in an infinite sound speed \citep{2010PhPl...17k2901B}. In addition, with the condition of incompressibility imposed, $p$ cannot be linked to other thermodynamic variables such as density and temperature through an equation of state, which causes the system to be thermodynamically inactive to any pressure perturbations \citep{kajishima2017}. Notably, as our primary goal is to analyze the topological changes that are responsible for the onset of flares, the assumed incompressibility seems to be justified in the rarefied coronal medium {\citep{dahlburg+1991apj}}.

The solution to the MHD equations as an initial value problem is derived from the well-established numerical model EULAG-MHD. This model is thoroughly detailed in \citep{2013JCoPh.236..608S} and the associated references. The advection scheme utilized in the model is the spatio-temporally second-order accurate non-oscillatory forward-in-time multi-dimensional positive definite advection transport algorithm known as MPDATA \citep{2006IJNMF..50.1123S}. A significant characteristic of MPDATA, which is vital for our simulation, is its dissipative property that is both intermittent and adaptive. This feature regularizes the under-resolved scales that arise in the magnetic field by simulating reconnections, mimicking the function of explicit subgrid scale turbulence models in the context of implicit large eddy simulation \citep[ILES,][]{grinstein2007book}. In our prior studies \citep{2018ApJ...860...96P,2019ApJ...875...10N,2020ApJ...903..129P,2021PhPl...28b4502N,2024ApJ...975..143N,2025SoPh..300...22K}, we have conducted extensive investigations into the dynamics of various active regions using this model and have successfully elucidated numerous events such as flares, coronal jets, and coronal dimming within these active regions.

\subsection{Simulation Setup}
The presented MHD simulation is carried out in a domain with computational grid points  $456\times 228 \times 228$ for a physical domain spanning $[0,2]\times [0,1]\times [0,1]$ units in $x$, $y$, and $z$, respectively, where a unit length approximately corresponds to $164$ Mm. 
The magnetic field obtained from the NFFF extrapolation (Figure \ref{fig2}) serves as an initial magnetic field for the simulation. The initial velocity field is taken to be zero. The initial Lorentz force (Figure \ref{fig3}(d)) pushes the plasma from the initial motion-less state and generates the dynamics. Further, with a minimal change in magnetic flux at the lower boundary during the flaring event, the vertical component of the magnetic field and velocity are fixed to their initial values at the boundary. The remaining components are linearly extrapolated from their interior points in their special neighbourhood \citep{2023A&A...677A..43P}. The constant $(\tau_{a}/\tau_{v})$ is set at $2\times10^{-4}$, which is 15 times greater than its coronal value, which speeds up the dynamical evolution without affecting the field topology. While satisfying the Courant-Friedrichs-Lewy (CFL) stability condition \citep{1967IBMJ...11..215C}, the time interval $\Delta {t}$ is set to $2\times10^{-3}$ (in the units of $\tau_a$). 
The Alfven transit time $\tau_{a}$ is 30 seconds. We perform the simulation for 2500 $\Delta{t}$. With the chosen $(\tau_{a}/\tau_{v})$, this roughly corresponds to 37 minutes of observation time. Here we present the time in units of $\tau_a$ while explaining the simulation results.

\section{Results of MHD Simulation}
\label{Results of MHD Simulation}
\subsection{Reconnection at HFT}
\begin{figure}
    \centering
    \includegraphics[width=1\linewidth]{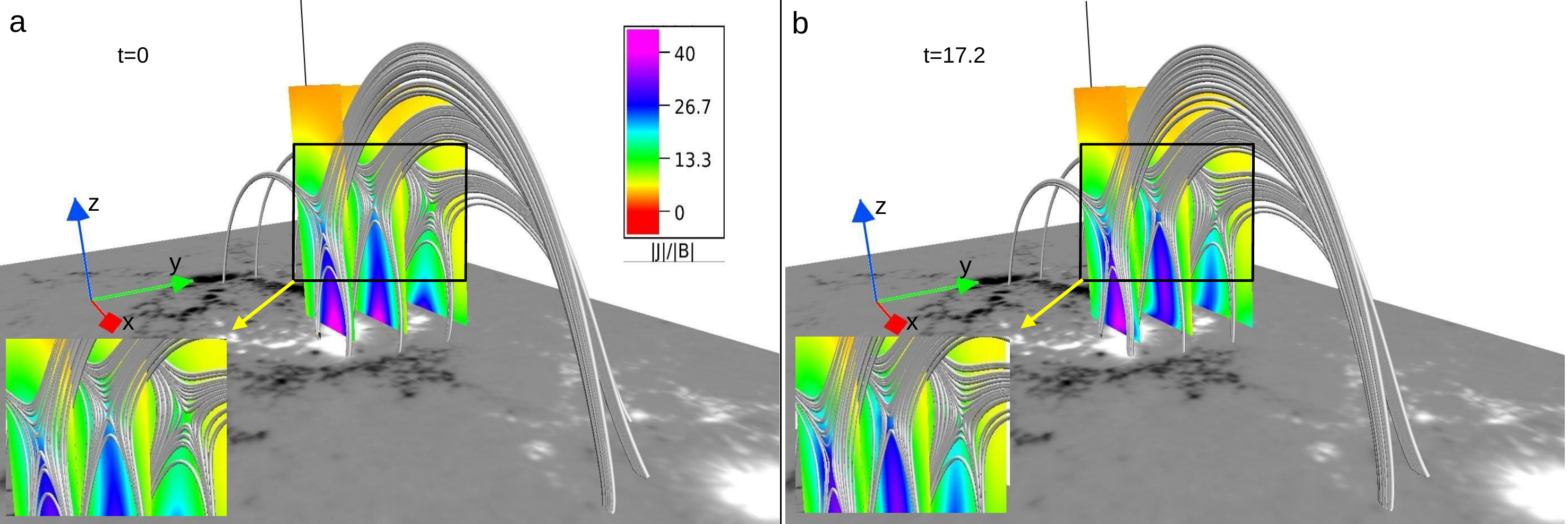}
    \caption{Plots of the projected field lines at three different $y$-constant planes located in the vicinity of the HFT at initial time $\text{t}=0$ (panel (a)) and $\text{t}=17.2$ (panel (b)). The plots are also overlaid with $|\textbf{J}|/|\textbf{B}|$. }
    \label{fig4}
\end{figure}

In the simulation, as mentioned above, the initial Lorentz force drives the active region's dynamics from a motionless state ($\textbf{v}=0$). To explore the possibility of magnetic reconnection at the HFT, in Figure \ref{fig4}, we plot the field lines of the transverse field in the vicinity of the HFT (as shown in Figure \ref{fig3}(c)) at $t=0$ and $t=17.2$. The figure is further overlaid with $|\textbf{J}|/|\textbf{B}|$. Notable is the enhancement in $|\textbf{J}|/|\textbf{B}|$ at $t=17.2$ in comparison to its initial values. Such enhancement corresponds to an increase in the gradient of magnetic field, resulted from the movement of the non-parallel magnetic field lines in close proximity {\citep{2025SoPh..300...79N}}. This also leads to the development of current sheets at the HFT, as suggested in the previous studies   \citep{2003ApJ...582.1172T,2003ApJ...595..506G,2005A&A...444..961A}. As the distance between the non-parallel field lines falls below the selected grid resolution, the scales become under-resolved, which are then regularized by magnetic reconnection in the simulation \citep{10.1063/1.4945634, Kumar_2016}.

\begin{figure}
    \centering
    \includegraphics[width=1\linewidth]{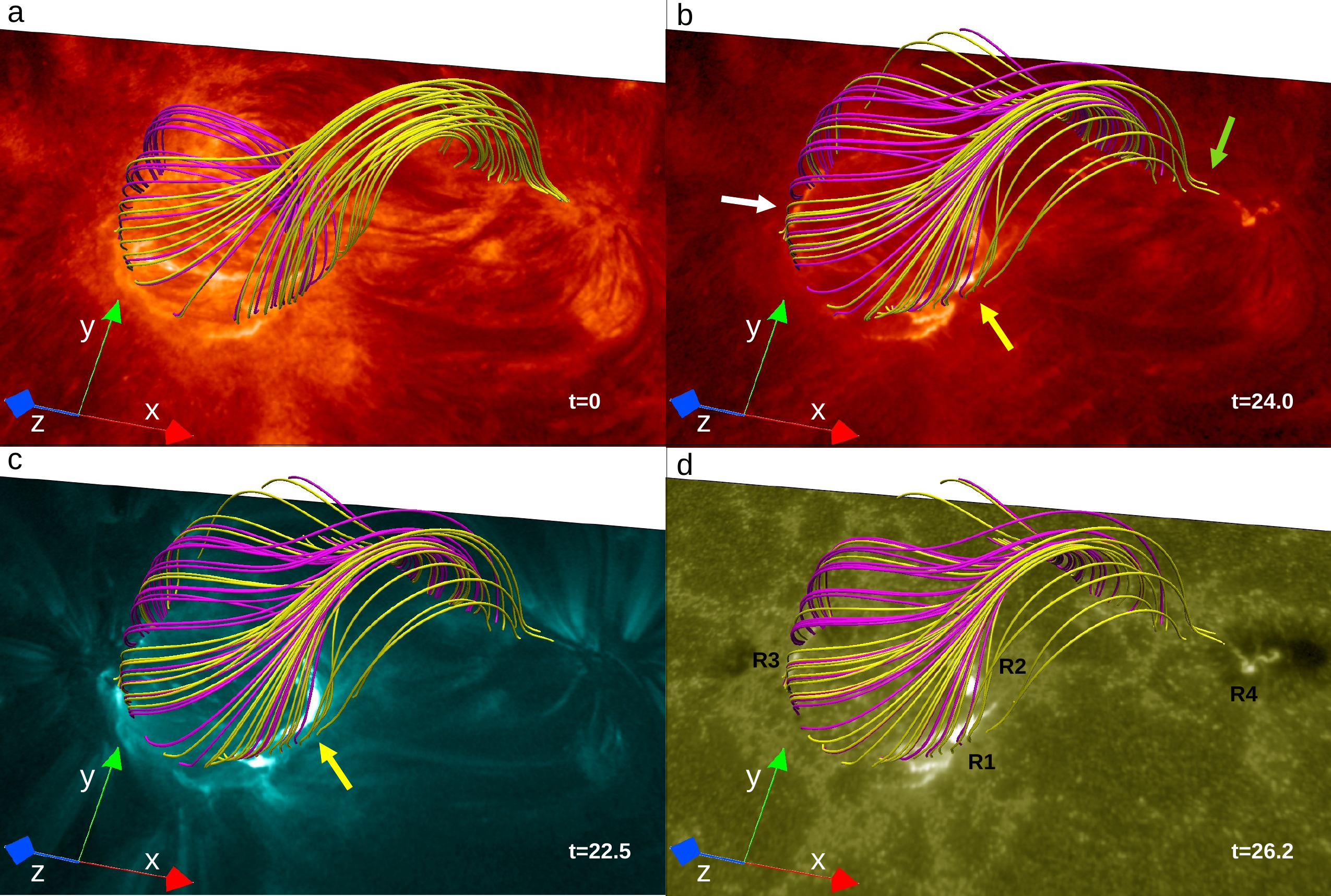}
    \caption{Simulated evolution of the magnetic field lines overlaid with the contemporary AIA 304 $\text{\AA}$ (panels (a) and (b)), 131 $\text{\AA}$ (panel (c)) and 1600  $\text{\AA}$ (panel (d)) images in the background. Yellow arrows denote the central brightenings, while white and green arrows highlight the remote brightenings, located left and right to the central brightening. The flare ribbons are identified by R1, R2, R3 and R4 in 1600  $\text{\AA}$.}
    \label{fig5}
\end{figure}

 To investigate the effects of the reconnection within the 3D magnetic configuration of the HFT, we present the time profile of the configuration in Figure \ref{fig5}, which is characterized by different connectivity represented by the pink and yellow field lines. Additionally, the figure includes contemporary AIA images across different wavelengths. Panels (a) and (b) depict the configuration alongside the contemporary AIA 304 $\text{\AA}$ images at times $t=0$ and $t=24$, respectively. It is important to highlight that initially, the pink field lines are connected between the negative polarity N and the positive polarity P2 (see Figures \ref{fig2} and \ref{fig5}(a)). As time progresses, these field lines start to connect between the negative polarity N and the positive plage polarity found to the east of P1 (refer to Figures \ref{fig2} and \ref{fig5}(b)) --- implying an alteration in the connectivity of the pink field lines.  This further corroborates the occurrence of magnetic reconnection at the HFT. Additionally, we note the emergence of central brightening beneath the HFT structure, as highlighted by a yellow arrow in panel (b). The charged particles accelerated from the reconnection site are expected to play a role in this brightening. Furthermore, the footpoints of the field lines in neighbourhood of the HFT closely align with the remote faint brightenings (denoted by white and green arrows in panel (b)) situated to the left and right of the central brightening. The charged particles originating from the reconnection site can travel along the field lines associated with the footpoints, potentially leading to these brightenings. Similar to the 304 $\text{\AA}$ observations, the central brightening beneath the HFT is also visible in the AIA 131 $\text{\AA}$ images (panel (c)), suggesting a causal relationship between the HFT reconnection and the brightening. In panel (d), we illustrate the field lines in the vicinity of the HFT overlaid on an AIA 1600 $\text{\AA}$ image at time $t=26.2$. Importantly, the ribbons R1 and R2 are positioned below the HFT, with the footpoints of the field lines nearly coinciding with the ribbon brightenings in the central region. Meanwhile, the other two remote ribbons, R3 and R4, are also almost co-located with the footpoints of the field lines close to the HFT on the left and right sides of the central region. This observation further implies the involvement of HFT reconnection in the formation of the flare ribbons.

\begin{figure}
    \centering
    \includegraphics[height=12cm, keepaspectratio]{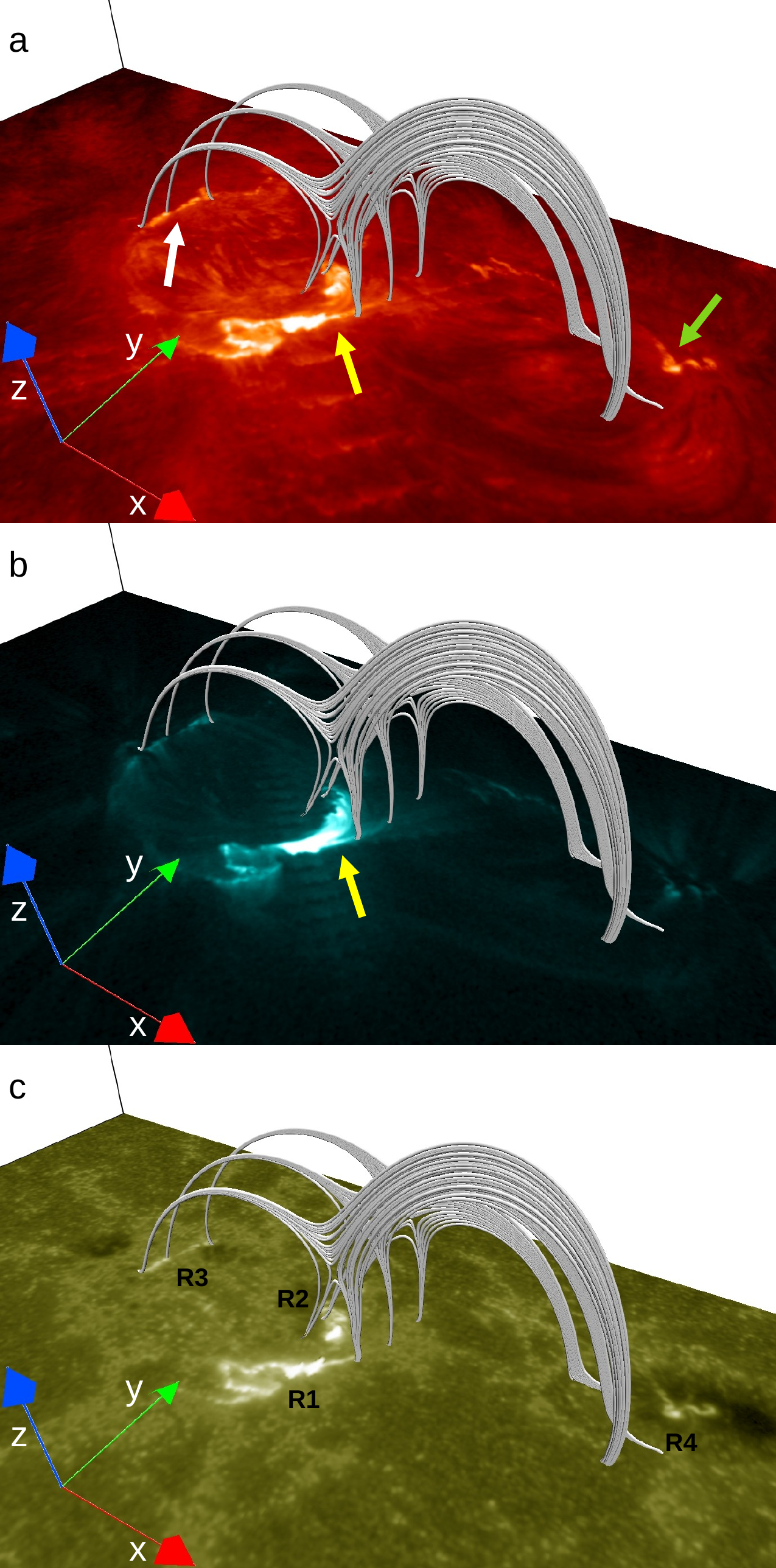}
    \caption{Plots of the projected field lines in different $y$-constant planes around the HFT at time $\mathbf{t=26.2}$ superimposed with 304 $\text{\AA}$ (panel (a)), 131 $\text{\AA}$ (panel (b)) and 1600 $\text{\AA}$ (panel (c)) images at the lower boundary. }
    \label{fig6}
\end{figure}

For a more clear correspondence of observed brightening and the field lines in neighbourhood of the HFT, in Figure \ref{fig6}, we present the X-shaped field lines of the 2D transverse field, plotted on three distinct $y$-constant planes at time $t=26.2$. Additionally, the figure features co-temporal AIA 304 $\text{\AA}$ (panel (a)), 131 $\text{\AA}$ (panel (b)), and 1600 $\text{\AA}$ (panel (c)) images at the lower boundary. In panels (a) and (b), it is evident that the central brightening (indicated by the yellow arrows) is almost co-spatial with the footpoints of the X-shaped field lines situated in the central area. The other two remote brightenings observed in the 304 $\text{\AA}$ image (highlighted by the white and green arrows in panel (a)) are located closer to the distant footpoints of the X-shaped field lines. Similarly, in panel (c), we observe that two quasi-parallel ribbons, R1 and R2, are co-located with the footpoints of the field lines in the central region, while two additional remote ribbons, R3 and R4, are closely linked to the far-situated footpoints. This further substantiates the involvement of HFT reconnection in the flare brightenings and the ribbons. In particular, the elongated and quasi-parallel nature of the R1 and R2 ribbons can be associated with the elongated intersection of two QSLs at the HFT (see Figure \ref{fig3}) that results in a series of X-shaped 2D field lines when projected in constant $y$ planes. Moreover, the early appearance of the central brightening and the primary ribbons (R1 and R2) can be attributed to their proximity to the reconnection site.  
The remote brightenings along with the secondary ribbons (R3 and R4) show a slight delay in their occurrence, which can be attributed to their greater distance from the reconnection site.

\subsection{Slipping Reconnection at the QSLs.}
\begin{figure}
     \centering
    \includegraphics[width=1\linewidth]{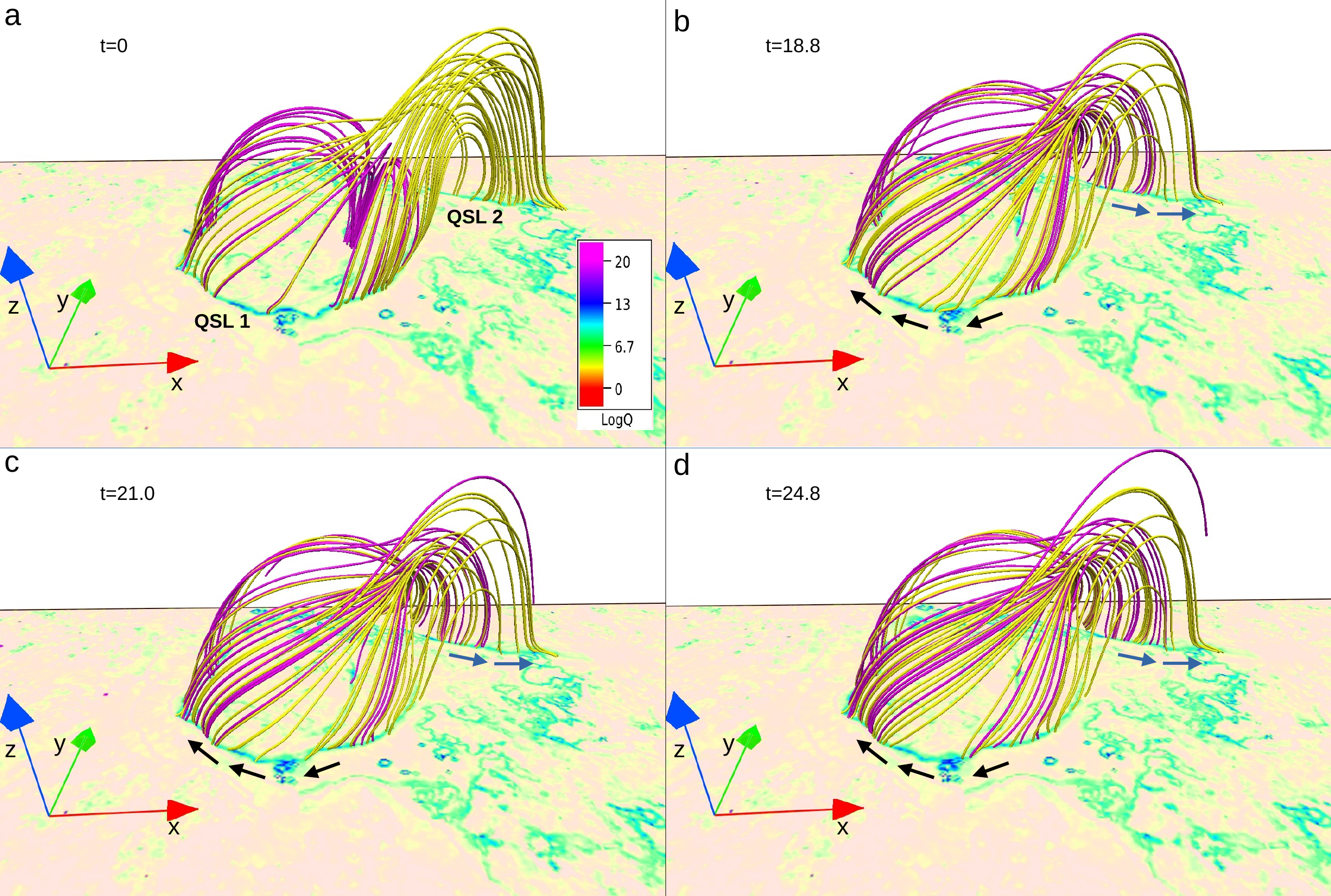}
    \caption{Time profile of magnetic field lines overplotted with LogQ at bottom boundary. QSL1 and QSL2 highlight the locations of two QSLs. Black and blue arrows (panels (b)-(d)) depict the direction of the motion of the field lines in QSL1 and QSL2, respectively.}
    \label{fig7}
\end{figure}

To assess the possibility of reconnection at the QSLs (as shown in Figure \ref{fig2}(d)), in Figure \ref{fig7}, we illustrate the time evolution of the field lines, overlaid with the squashing factor $Q$ at the bottom boundary. The QSLs are characterized as regions with high Q-values (LogQ $>$ 8), labeled as QLS1 and QSL2 in panel (a). The movement of the footpoints of the field lines in QSL1 and QSL2 is represented by black and blue arrows in panels (b)-(d), respectively. It is observable that the movement is such that the footpoints primarily lie on the QSLs, indicating towards the onset of slipping magnetic reconnections at these QSLs \citep{2006SoPh..238..347A}. We have also investigated the direction of plasma flow in the QSLs, which is distinct from the motion of the footpoints (not shown), thereby providing further evidence for the occurrence of slipping reconnection.

\begin{figure}
    \centering
    \includegraphics[width=1\linewidth]{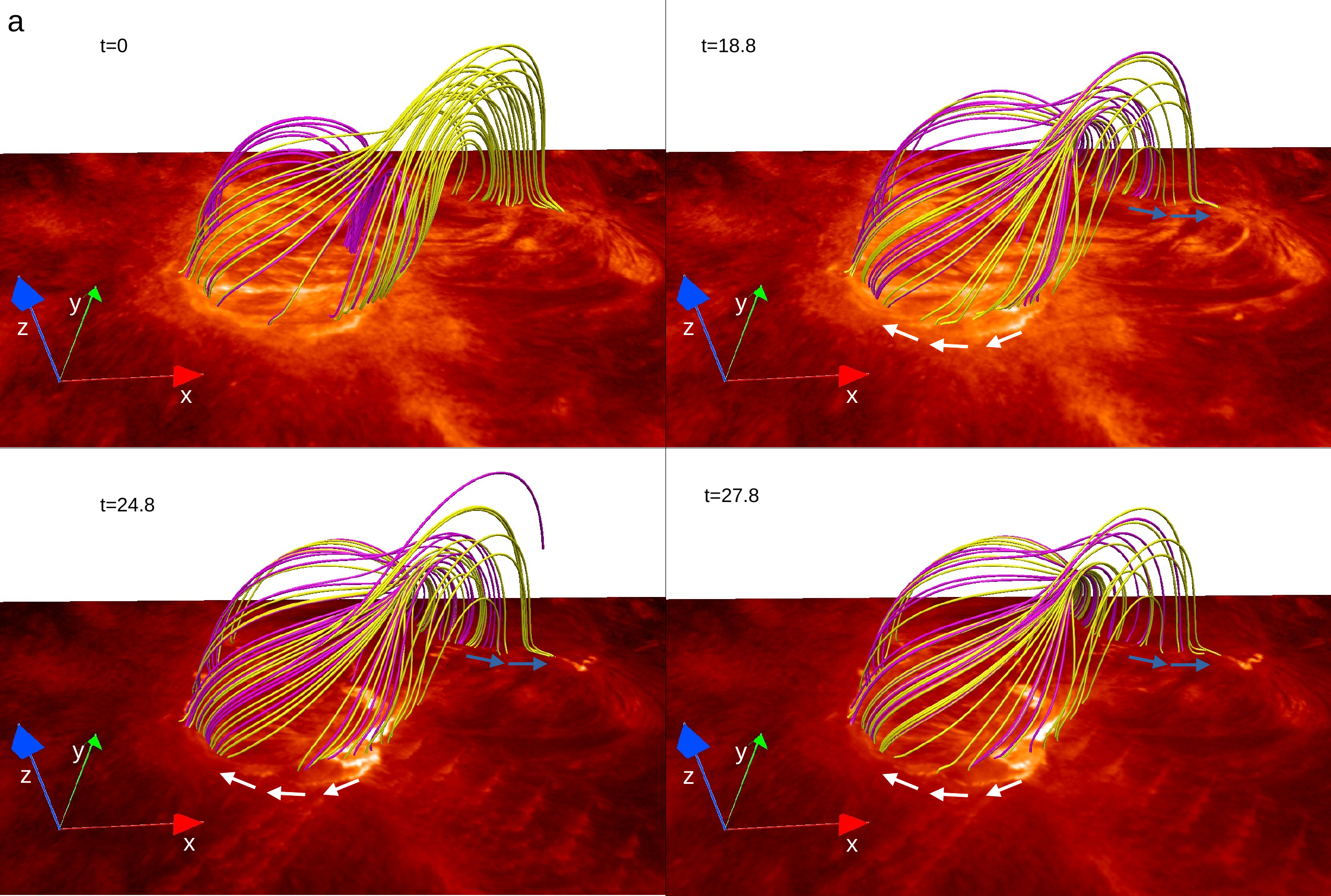}
    \caption{The figure illustrates the magnetic field line evolution overlaid with co-temporal 304 $\text{\AA}$ images. White and blue arrows (panels (b)-(d)) denote the movement of the footpoints of the field lines.}
    \label{fig8}
\end{figure}

To explore the correspondence between the slipping reconnection with the flare brightenings, 
in Figure \ref{fig8}, the simulated evolution of the magnetic field lines is overlaid onto the co-temporal observations in 304 $\text{\AA}$.  Importantly, the movement of the field lines in QSL1
(marked in white arrows in Figure \ref{fig8}(b)-(d)) exhibits similar directionality as the clockwise extension of the central brightening in the form of the faint quasi-circular brightening (as illustrated in Figure \ref{fig1}(d)). This indicates that the
underlying slipping reconnections are responsible for the generation of the quasi-circular faint brightening in this region. Furthermore, the motion of field lines in QSL2  (marked by blue arrows in Figure \ref{fig8}(b)-(d)) is approximately co-align with the faint brightening trace (as shown in the white rectangle in Figure \ref{fig1}(c)) --- suggesting the role of the slipping reconnection in the faint brightening. 

\section{Summary and Discussion}
\label{Summary and Discussion}
In this paper, we explore the onset of a confined M2.1 class flare which is initiated at approximately 11:32 UT in NOAA active region 12268 on 2015 January 29. using a data-constrained MHD simulation. The simulation was initiated through a novel non-force-free-field extrapolation. 
The NFFF extrapolated field is generated from the HMI/SDO photospheric vector magnetogram of the active region, taken roughly 8 minutes prior to the flare's initiation.

Multi-wavelength observations of the flaring event document a brightening in the central portion of the flaring region, along with two remote brightenings located to the left and right of the central brightening (see Figures \ref{fig1}). As time progresses, we observe a clockwise expansion of the central brightening, which leads to the formation of a faint quasi-circular brightening structure (refer to panels (d) and (f) of Figure \ref{fig1}). Moreover, it is noteworthy that a faint brightening trace is generated (illustrated in Figure \ref{fig1}(c)). Observations in 1600 $\text{\AA}$ show the emergence of two primary quasi-parallel flare ribbons, designated R1 and R2 (as shown in Figure \ref{fig1}(g)). Subsequently, two additional ribbons, R3 and R4, are observed to develop (Figure \ref{fig1}(h)), which are fainter in comparison to R1 and R2.

The initial extrapolated magnetic field shows an intriguing magnetic topology in the flaring region. The configuration of the overall magnetic field lines reveals two distinct connectivity domains that resemble the fan-spine structure of a three-dimensional null (see Figure \ref{fig2}(b)-(c)).  However, the trilinear null detection technique did not reveal any null. Importantly, we discover an intersection of two QSLs at some heights in the computational domain, where the squashing factor at this intersection is considerably higher than in adjacent regions (Figure \ref{fig3}(a)-(c)). Noticeably, the higher squashing factor values are arranged in an elongated configuration. The projected 2D field lines exhibit an X-shaped geometry around the intersection points, suggesting the existence of a HFT (Figure \ref{fig3}(b)). Additionally, we identify two QSLs at the bottom boundary of the domain (as depicted in Figure \ref{fig2}(d)).

The simulated dynamics reveals a significant enhancement in  $|\textbf{J}|/|\textbf{B}|$ at the HFT (Figure \ref{fig4}), which suggests the development of the current sheet at the HFT. This leads to the commencement of magnetic reconnection at the HFT. It is proposed that the charged particles accelerated from the reconnection site are involved in the pre-flare brightenings (including both the central brightening and the remote brightenings) as well as in the generation of the flare ribbons. The delayed appearance of the remote brightenings and the secondary ribbons (R3 and R4) is attributed to their positioning far from the reconnection site compared to the central brightening and the primary ribbons (R1 and R2).
The elongated and quasi-parallel arrangement of ribbons R1 and R2 is thought to be associated with the elongated intersection of two QSLs at the HFT. Furthermore, the occurrence of slipping reconnection at the QSLs at the bottom boundary is significant. The slipping reconnection in QSL1 is suggested to result in the clockwise extension of the central brightening, forming the quasi-circular faint bright structure. Moreover, the faint brightening trace is attributed to the slipping reconnection in QSL2.

Overall, the data-constrained MHD simulation effectively illustrates the significance of magnetic reconnection in complex 3D magnetic structures, including HFT and QSLs, in the onset of a confined flare, thus enhancing our comprehension of flare initiation. Notably, prior investigations of confined flares characterized by circular or quasi-circular brightenings and corresponding ribbons indicated the pivotal role of magnetic reconnection at the 3D null points in the commencement of these flares \citep{2009ApJ...700..559M, 2018ApJ...860...96P, 2025SoPh..300...22K}. In contrast, the current work emphasizes the role of magnetic reconnection at HFT and QSLs in such a flare. In the future, we plan to enhance the realism of the simulation by relaxing the incompressibility condition and incorporating an appropriate physical resistivity.

\vspace{1cm}

\textbf{Acknowledgments} We acknowledge Dr. Avijeet Prasad for his valuable suggestions to enhance the quality of the work. We acknowledge the use of the visualization software VAPOR (www.vapor.ucar.edu) for generating relevant graphics. The MHD simulation was performed on the Param Vikram-1000 High Performance Computing Cluster of the Physical Research Laboratory (PRL). Part of the computation was carried out on the computing cluster Pegasus of IUCAA, Pune, India. Data and images are courtesy of NASA/SDO and the HMI and AIA science teams. SDO/HMI is a joint effort of many teams and individuals to whom we are greatly indebted for providing the data. This work is supported by the Department of Space, Government of India. We are thankful to the referee for providing thoughtful suggestions and comments, which have greatly improved this paper.

\vspace{0.5cm}

\textbf{Author contributions} P.K., S. and S.K. analyze the observations and simulation data. P.K. and S. prepared the plots. P.K. and S.K. wrote the main draft of the paper. Simrat helped in setting-up the MHD simulation. S.S.N. and R.B. contributed to the interpretation of the results. All authors did a careful proofreading of the text and improved the quality of the paper.

\vspace{0.5cm}

\textbf{Fundings} S.K., S. would like to acknowledge the support from the ANRF-SERB project No. SUR/2022/000569. S.K. also acknowledges support from Patna University project No. RDC/MRP/07. S.S.N. acknowledges the supports from the I+D+I project PID2023-147708NB-I00 funded by FEDER,EU and MICIU/AEI/10.13039/501100011033/.

\vspace{0.5cm}

\textbf{Data Availability} The data that support the findings of this study are available from the
corresponding author upon reasonable request.

\vspace{0.5cm}

\textbf{Conflict of interest} The authors declare that they have no conflicts of interest.

\bibliographystyle{spr-mp-sola}
\bibliography{ref}
\end{document}